\documentclass[epj,referee]{svjour1}

\usepackage{graphics}

\begin{document}

\title{Polar modes in multiferroic $Ba_2Mg_2Fe_{12}O_{22}$ – hexagonal Y-type hexaferrite}
\author{G.A. Komandin\inst{1} \and A.S. Prokhorov\inst{1,2} \and V.I. Torgashev\inst{3} \and A.A. Bush\inst{4} \and E.S. Zhukova\inst{1,2} \and
B.P. Gorshunov\inst{1,2,}\thanks {E-mail: gorshunov@ran.gpi.ru} }
\institute{A.M.Prokhorov Institute of General Physics, Russian Academy of Sciences, Vavilov str. 38, 119991 Moscow, Russia \and
Moscow Institute of Physics and Technology (State University), Institutskii lane 9, 141700, Dolgoprudnyi, Russia \and
Faculty of Physics, Southern Federal University, Zorge Str. 5, 344090 Rostov-on-Don, Russia \and
Moscow Institute of Radiotechnics, Electronics and Automation, 117464, Moscow, Russia}

\abstract{ Spectra of transmission and reflection coefficients of
single crystalline $Y$-type hexaferrite \\ $Ba_2Mg_2Fe_{12}O_{22}$ are
measured at Terahertz and infrared frequencies, $3\ -\ 4500\
cm^{-1}$, and at temperatures 10 to 300 K, for polarization
$\vec{E}\perp c$. Oscillator parameters of nineteen phonon lines
allowed by the $R3m$ crystal lattice symmetry are determined.
Absorption lines assignment to vibrational modes of the lattice
structural fragments is performed. Below the temperatures of $195\
K$ and $50\ K$, corresponding to the zero field phase transitions
to the proper screw and to the longitudinal-conical spin states,
respectively, new absorption lines are discovered whose origin is
assigned to the electric-dipole active magnetic excitations.
 }

\maketitle

\section{Introduction}
\label{intro}
\par After the discovery of a huge magneto-electric effect in $TbMnO_3$ \cite{1}, much attention is being paid to a class of ferroelectrics with cycloidal spin order \cite{2,3}. In such materials the application of relatively small magnetic field (few Tesla) can induce various phases including those revealing ferroelectricity. Model objects for studying magnetic and ferroelectric effects are hexaferrites, typical representatives are compounds of the M-type, like $AFe_{12}O_{19}$, $Y$-type, like $A_{2}B_{2}Fe_{12}O_{22}$, W-type, like $AMe_{2}Fe_{16}O_{27}$, Z-type, like $A_3Me_2Fe_{24}O_{41}$, X-type, like $A_2Me_2Fe_{28}O_{46}$, and U-type, like $A_4Me_2Fe_{36}O_{60}$; here $A$ = $Sr$, $Pb$, $Ba$; $M$ = $Zn$, $Fe$, $Co$, $Mg$, $Mn$ \cite{4}. Of special interest are $Y$-type materials in which unique multiferroic properties have recently been discovered \cite{5,6,7,8,9,10,11,12,13,14,15,16}. In particular, in $Ba_2Mg_2Fe_{12}O_{22}$ ($BMFO$), magnetoelectric switching can be realized by application of rather small magnetic field \vec{B}  (the value of \vec{B} can be as low as $\pm30\ mT$ \cite{6}), while $Ba_{0.5}Sr_{1.5}Zn_2(Fe_{1-x}Al_x)_{12}O_{22}$ can be efficiently tailored by the $Al$-substitution (at $x = 0.08$ the critical magnetic field for switching the electric polarization is reduced down to $\sim 1\ mT$ \cite{12}). The phase diagrams of these compounds contain several magnetically ordered phases and some of them are also ferroelectric. The crystal structure of $Y$-type hexaferrites is rather complicated. For instance, the unit cell of $Ba_2M_2Fe_{12}O_{22}$ ($M$ = $Zn$, $Fe$, $Co$, $Mg$, $Mn$) contains 18 atomic layers with $(cchhhh)_3$ stacking sequence along the $c$-direction \cite{17,18}, as indicated in figure
1.
\par Magnetic structure of Y-hexaferrites is under intensive study for already more than 50 years \cite{13,15,16,19,20,21,22,23,24}. In \cite{19}, an existence of a collinear-spin ferrimagnetic order was suggested. Later, an idea about helical spin ordering in $(Sr,Ba)_{2}Zn_{2}Fe_{12}O_{22}$ was put forward \cite{20} and experimentally confirmed in \cite{15,16,21,22,23,24}. According to \cite{15}, the magnetic structure in $Ba_2Mg_2Fe_{12}O_{22}$ is proper screw below $195\ K$ at $\vec{B} = 0$ (figure 1b) and collinear at higher temperatures (figure 1a). Application of magnetic field can generate a series of magnetic phase transitions \cite{6,7,10,11,13}. In $Ba_2Mg_2Fe_{12}O_{22}$ there exists a longitudinal conical spin ordered state below $50\ K$ (figure 1c) \cite{13}. Here, ferroelectric phases with a quasitransverse spin ordering \cite{6,13} (figure 1d) can be induced by applying magnetic field of magnitude much smaller than in $Ba_{0.5}Sr_{1.5}Zn_2Fe_{12}O_{22}$ \cite{5,8}. It turned out that the phase boundaries can be effectively tuned by heat treating \cite{8} as well as by $Al$-substitution into octahedral Fe sublattices \cite{12}. It should be noted that recent neutron diffraction studies \cite{10,13} revealed much more complicated magnetic phase diagram of $Ba_2Mg_2Fe_{12}O_{22}$ than was reported previously
\cite{6}.

\begin{figure}
\centering
\resizebox{0.6\columnwidth}{!}
%\vspace{17cm}
{\includegraphics{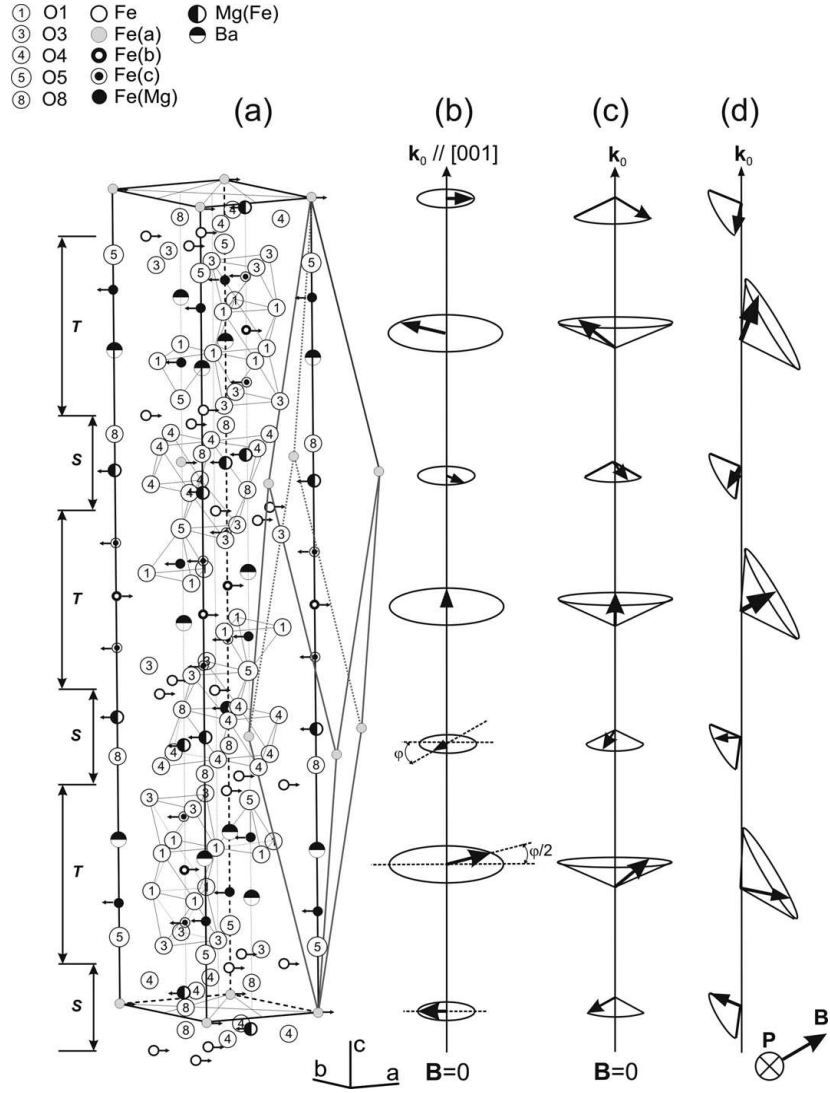} }
\caption{(a) -
Unit (hexagonal) and primitive (rhombohedral) cells of
$Ba_2Mg_2Fe_{12}O_{22}$ hexaferrite. Coordinating polyhedrons of
some cations in different cell sites are shown. $T$ and $S$ stand
for two kinds of magnetic sublattices with large and small
magnetic moments. (b) - (d) Schematic illustration of helix spin
ordering in different magnetically ordered phases. (b), (c) and
(d) correspond to the proper screw $(50 < T < 195\ K)$,
longitudinal conical $(T < 50\ K)$ and quasitransverse $(T < 195\
K, \vec{B}\neq0)$ spin orderings, respectively. Spins orientation
in the ferrimagnetic phase is indicated by arrows.} \label{fig:1}
\end{figure}

\par Since hexaferrites with magnetically ordered long periodic structures reveal strong magnetoelectric effects at elevated temperatures (above $77\ K$) and in relatively small magnetic fields, they are promising for application in magnetoelectric systems. Presently, the mechanisms of competition between magnetic and electric interactions in hexaferrites remain unclear. That is why the aim of this work was to further study the mechanisms of formation of magnetic and ferroelectric phases in $Ba_{2}Mg_{2}Fe_{12}O_{22}$ by using the Terahertz-infrared (THz-IR) spectroscopy. Indeed, finite interaction between the two subsystems, spins and phonons, would necessarily lead to a noticeable transformation in the phonon subsystem (IR range) in the materials and these changes can be used for getting more information about magnetoelectric properties of the compounds. Besides, recently, Kida et al. have reported on the observation in $Ba_2Mg_2Fe_{12}O_{22}$ of electric-dipole active magnetic excitations \cite{11}, called electromagnons. The excitations were detected at THz frequencies in longitudinal-conical spin state and revealed huge changes in the intensity under application of magnetic field less than $0.2\ T$ (along [001]). On the other hand, Kamba et al. report \cite{14} that in $Ba_{0.5}Sr_{1.5}Zn_2Fe_{12}O_{22}$ \textit{"no evidence for ferroelectricity was observed either on cooling (at \vec{B}=0) or after application of external magnetic field up to $0.3\ T$ at $300\ K$"}. To understand the reason of these discrepancies was also one of our
aims.

\section{Experimental details}
\label{Experiment}
\par Single crystalline samples of $Ba_2Mg_2Fe_{12}O_{22}$ were grown by a flux method as described in \cite{16}. For the infrared and THz measurements plane-parallel samples were prepared of thickness about $0.5\ mm$ and size about $5\times5\ mm^2$, with the $c$ axis perpendicular to the surface of the samples. In this geometry, only the $E_u$ modes can be registered by optical measurements, corresponding to the situation when the \vec{E}-vector of the probing radiation is perpendicular to the $c$-axis. The measurements were performed on two spectrometers. For the low frequency range, $15 - 35\ cm^{-1}$, a quasioptical THz spectrometer based on backward-wave oscillators was used. With this instrument by measuring complex transmission coefficient spectra $Tr(\nu)$, the values of real $\epsilon'$ and imaginary $\epsilon''$ parts of dielectric permittivity are directly obtained \cite{25}. A Fourier-transform spectrometer Bruker IFS 113V was used to cover higher frequency range up to $4500\ cm^{-1}$. At these frequencies the reflection coefficient spectra $R(\nu)$ were measured. The spectra obtained on the two spectrometers, $Tr(\nu)$, $R(\nu)$, $\epsilon'(\nu)$ and $\epsilon''(\nu)$, were merged and analyzed by a least square fitting procedure. The absorption lines were modeled with Lorentzians and coupled Lorentzians with the complex dielectric permittivity given by the following expressions. For
Lorentzians
\begin{equation}
    \label{1}
    \epsilon(\nu)=\epsilon_{\infty}+\sum _{j=1}^{n}{\frac {\Delta \epsilon_j\,{\nu_{{j}}}^{2}}{{\nu_{{j}}}^{2}-{\nu}^{2}+i\nu \,\gamma_{{j}}}}
\end{equation}
where $\epsilon_{\infty}$ is the high frequency dielectric
constant, $\Delta\epsilon_j$ is the dielectric contribution of the
absorption line, $\nu_j$ is its eigenfrequency and $\gamma_j$ the
damping constant. For the coupled Lorentzians the corresponding
expression is written as \cite{26}:

\begin{equation}
    \label{2}
    \epsilon  \left( \nu \right) ={\frac {s_{{2}} \left( {\nu_{{2}}}^{2}-{\nu}^{2}+i\nu \gamma _{{2}} \right) +s_{{2}} \left( {\nu_{{1}}}^{2}-{
\nu}^{2}+i\nu \gamma _{{1}} \right) -2\,\sqrt {s_{{1}}s_{{2}}} \left( \alpha+i\nu \delta  \right) }{ \left( {\nu_{{1}}}^{2}-{\nu}^{2}+i\nu \gamma _{{1}} \right)  \left( {\nu_{{2}}}^{2}-{\nu}^{2}+i\nu \gamma _{{2}} \right)-\left( \alpha+i\nu \delta  \right) ^{2}}}
\end{equation}
where $j$ = 1, 2, $s_j = \Delta\epsilon_j\nu_j^2$ is the
oscillator strength of $j$-th Lorentzian with the eigenfrequency
$\nu_j$; $\alpha$ is the real and $\delta$ – the imaginary
coupling constants.

\section{Results and discussion}
\label{sec:Result}

Examples of the spectra of the $Ba_2Mg_2Fe_{12}O_{22}$ sample
obtained at 10 K are presented in figure 2. Dots show the spectra
got on the THz and Fourier spectrometers and the lines represent
the results of the least-square processing of the spectra based on
the expressions \ref{1} and \ref{2}. At the lowest frequencies, an
oscillating behavior in the spectra is observed that comes from
interference of the radiation within the plane-parallel samples.
Processing the oscillations allows to determine directly the
dielectric parameters of the material, the real and the imaginary
parts of the dielectric constant, of the refractive index, etc.
\cite{25}. Figure 3 shows examples of the $\epsilon'$ and
$\epsilon''$ spectra of $Ba_2Mg_2Fe_{12}O_{22}$. Dots correspond
to the values of $\epsilon'$ and $\epsilon''$ directly obtained on
the THz spectrometer \cite{25} while lines show the results of the
fitting of the infrared spectra. Parameters of the absorption
lines are summarized in \ref{tab:1}.

\subsection{Spectral analysis}
\label{sec:Spectral}
\par We have performed a factor-group analysis of possible phonon modes in $Ba_2Mg_2Fe_{12}O_{22}$, with the results summarized in table 2. According to this table, 34 modes can be observed in the IR-spectra corresponding to the irreducible representations $15A_{2u}\ [IR: z]+19E_u\ [IR: x,y]$ of the space group $R\bar{3}m$. Nondegenerate $A_{2u}$-modes can be excited by electric field $ \vec{E} \parallel c$ and doubly degenerate $E_u$-modes - by the $\vec{E}$-field polarized perpendicular to the $c$-axis. To identify the excitations observed in our experiment, we compare spectra of $Ba_2Mg_2Fe_{12}O_{22}$ with the ones of magnetoplumbite and spinels \cite{27,28,29,30,31,32,33}. We consider two types of tetrahedra (M6 and M9 in table 2). An ideal $FeO_4$ tetrahedron with the $T_d$ symmetry has four fundamental vibration modes \cite{34}: $\nu_1(A_1)$, $\nu_2(E)$, $\nu_3(F_2)$, and $\nu_4 (F_2)$, of which only $\nu_3$ and $\nu_4$ are
IR-active. Due to the low local symmetry of
$Ba_2Mg_2Fe_{12}O_{22}$, also the $\nu_{2}$ modes are active in
optical spectra. We assume that the lines seen at $539$ and $580\
cm^{-1}$ are caused by the stretching $\nu_3$ vibrations and that
the deformational $\nu_4$ modes of the tetrahedra $(Fe/Mg)O_4$ and
$(Mg/Fe)O_4$ are located within a complicated absorption structure
in the range around $419\ cm^{-1}$. The modes $\nu_{2}$ have small
dipole moment and should be observable in the range $300-350\
cm^{-1}$. The $\nu_1$ mode is not IR-active. The complicated shape
of the $\nu_3$ and $\nu_4$ bands can be caused by a LO-TO
splitting of the polar modes or by a certain disorder of cations
($Fe/Mg$ and $Mg/Fe$) at the $(6c)$ sites within the tetrahedron
leading to different eigenfrequencies.

\begin{figure}
\centering
\resizebox{0.55\columnwidth}{!}
%\vspace{6cm}
{
 \includegraphics{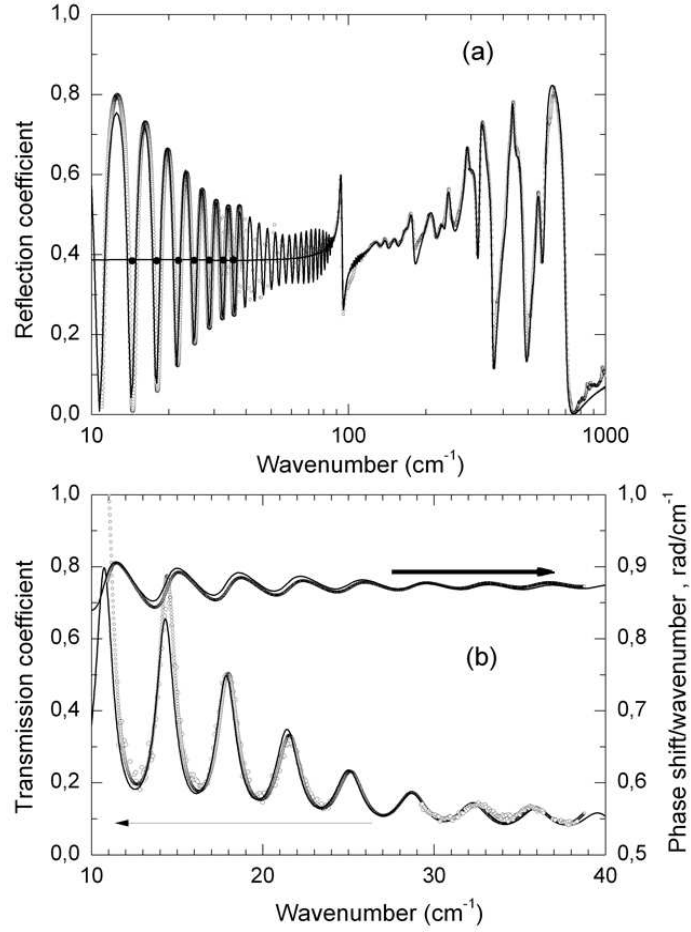}
}
 \caption{Spectra (dots)
of reflection (a) and complex transmission (amplitude and phase
shift) (b) \cite{25} of the $Ba_2Mg_2Fe_{12}O_{22}$ sample
measured at $T = 10\ K$ for polarization $\vec{E}\perp c$, and
their model processing (lines, see text). Filled symbols in panel
(a) are obtained by processing the THz transmission coefficient
and phase shift spectra of the type shown in panel (b).}
\label{fig:2}
\end{figure}

\begin{figure}
\centering
\resizebox{0.55\columnwidth}{!}{
%\vspace{6cm}
\includegraphics{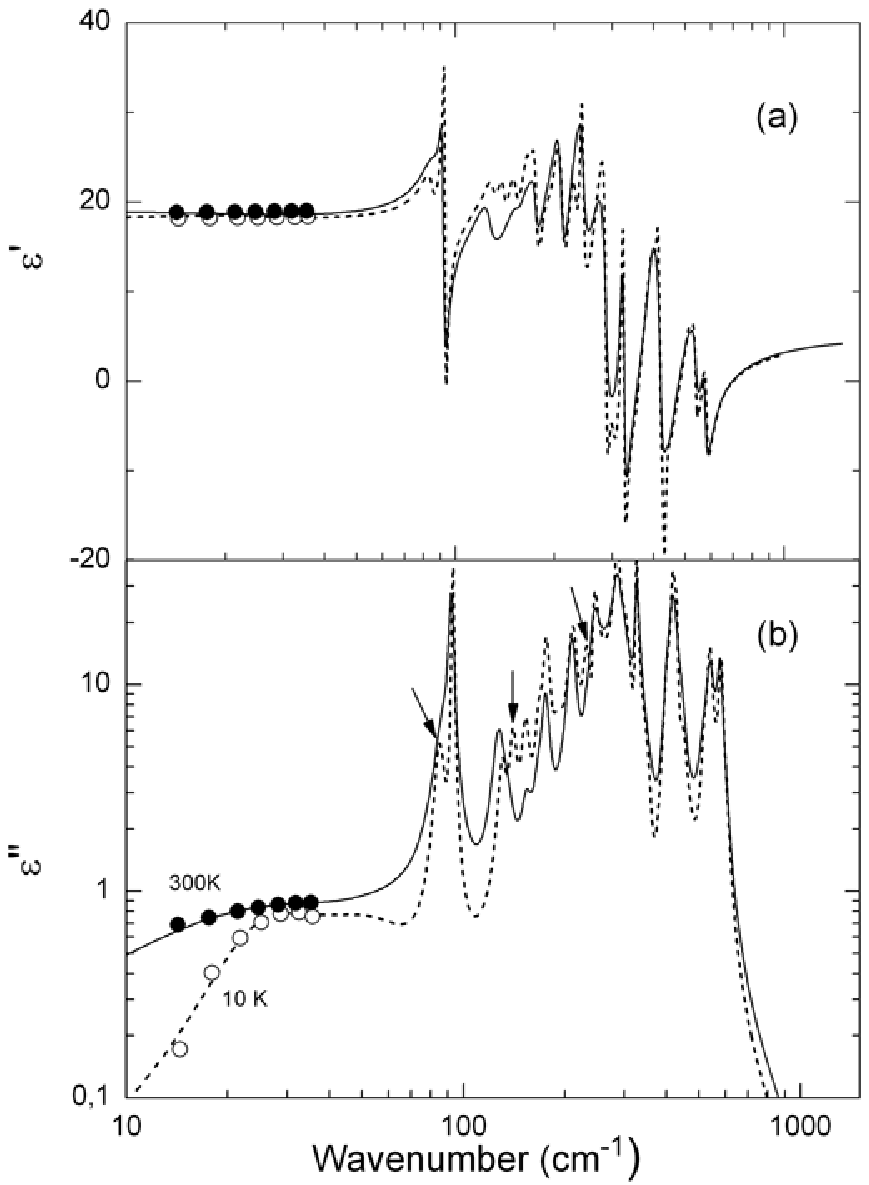}
} \caption{Spectra of real (a) and imaginary (b) parts of
dielectric permittivity of $Ba_2Mg_2Fe_{12}O_{22}$ single crystal
at $300\ K$ and $10\ K$, for polarization $\vec{E}\perp c$. Dots
at low frequencies (below 40 cm$^{-1}$) show the data obtained
from transmission coefficient spectra of a plane-parallel sample,
see text.} \label{fig:3}
\end{figure}

\par An undistorted $FeO_6$ octahedron has a $O_h$ symmetry and creates 6 fundamental modes \cite{34}: $\nu_1(A_{1g})$, $\nu_2(E_g)$, $\nu_3(F_{1u})$, $\nu_4(F_{1u})$, $\nu_5(F_{2g})$, and $\nu_6(F_{2u})$. Vibrations $\nu_1$, $\nu_2$, and $\nu_5$ should be Raman active, $\nu_3$ and $\nu_4$ - IR-active and $\nu_6$ is expected to be silent. Since the $FeO_6$ octahedra form chains and thus are not independent, unambiguous assignment of the observed absorption bands to octahedral $\nu_1-\nu_6$ modes is problematic. That is why only tentative assignment of these bands is given in table 1. Still, it is reasonable to assume that the bands at $240-300\ cm^{-1}$, $350-400\ cm^{-1}$ and $442\ ñm^{-1}$ correspond to the $\nu_3$ and $\nu_4$ vibrations of the octahedra. The bands at $150-210\ cm^{-1}$ are mainly coming from the bending vibrations of the concatenated $O-Fe-O-Fe-O$ bonds of adjoint octahedra and could also be assigned to the $\nu_3$ and $\nu_4$ molecular vibrations of the octahedron. In \cite{30} it is assumed that the absorption bands located close to $150\ cm^{-1}$ and $175\ cm^{-1}$ in the spectra of $M$-type hexaferrite $BaFe_{12}O_{19}$ are caused by vibrations of the bipyramid. The structure of $Ba_2Mg_2Fe_{12}O_{22}$, however, does not contain bipyramids and thus our assignment of the bands in this range to the bending vibrations looks more correct. Finally, the line around $92\ cm^{-1}$ should be assigned to the $Ba-O$ bonds vibration. We note that in $BaFe_{12}O_{19}$ this vibration has slightly larger frequency, of $100-102\ cm^{-1}$ and $123-125\ cm^{-1}$, according to
\cite{28,30,31}.

\begin{table}
\caption{Parameters of modes detected in the spectra of single
crystalline $Ba_2Mg_2Fe_{12}O_{22}$ for polarization $\vec{E}\perp
c$ at $T=10\ K$: frequency position $\nu_j$, damping $\gamma_j$,
dielectric contribution $\Delta\epsilon_j$; $\alpha$ and $\delta$
characterize the modes coupling between the modes. The last column
shows assignments of the modes to the type of displacement of the
structure-forming atoms. The high-frequency dielectric
permittivity is $\epsilon_\infty=4.52$.} \label{tab:1}
\begin{tabular}{|l|l|l|l|l|l|l|}
\hline\noalign{\smallskip}
Oscillator type & $\nu_j, cm^{-1}$ & $\gamma_j\ ,cm^{-1}$ & $\Delta\epsilon_j$ & $\alpha$ & $\delta$ & Assignment, ions involved   \\
\noalign{\smallskip}\hline\noalign{\smallskip}
coupled Lorentzians 1 & 29.08 & 20.52   & 0.36 &&& magnon-phonon continuum \\
\cline{2-6}
             & 56.78 & 41.32 &  0.36    &&  -0.86 & \\
\hline
Lorentzian 2 & 93.21 & 1.62 &   0.67    & 0  && $Ba-O$ \\
\hline
Lorentzian 3 & 130.41 & 8.27 &  0.20 &&& $Fe-O-Fe,\ (FeO_6)$ \\
\hline
Lorentzian 4 & 140.65 & 7.41 &  0.23 &&& optic magnon \\
\hline
Lorentzian 5 & 152.93 & 9.11 & 0.30 &&& $Fe-O-Fe,\ (FeO_6)$ \\
\hline
Lorentzian 6 & 167.45 & 9.11 &  0.21 &&& $Fe-O-Fe,\ (FeO_6)$ \\
\hline
Lorentzian 7 & 176.29 & 7.96 & 0.75 &&& $Fe-O-Fe,\ (FeO_6)$ \\
\hline
Lorentzian 8 & 211.54 & 17.54   & 1.40 &&& $Fe-O-Fe,\ (FeO_6)$ \\
\hline
Lorentzian 9 & 233.52 & 19.55   & 1.21  &&& optic magnon \\
\hline
coupled Lorentzians 10 & 243.14 & 8.93 &    0.45 &  0   && $Fe-O,\ (FeO_6)$, bend \\
\cline{2-7}
& 288.47 & 11.10 &  1.13 && -43.38 & $Fe-O,\ (FeO_6)$, bend \\
\hline
coupled Lorentzians 11 & 301.30 &   32.57 & 2.50 & 0 && $Fe-O,\ (FeO_6)$, bend \\
\cline{2-7}
& 325.10 & 6.54 & 0.87 && 6.73 & $\nu_2\ (FeO_4)$, bend \\
\hline
coupled Lorentzians 12 & 417.59 &   13.21 & 0.75 & 0    && $\nu_4\ (Fe/Mg)O_4$, bend \\
\cline{2-7}
& 431.33 & 11.58 & 0.12 && -13.11 & $\nu_4\ (Mg/Fe)O_4$, bend \\
\hline
coupled Lorentzians 13 & 353.49 & 28.33 &   0.09 & 0 && $Fe-O,\ (FeO_6)$, bend \\
\cline{2-7}
& 403.86    & 29.65 & 0.80 && 38.23 & $Fe-O,\ (FeO_6)$, bend \\
\hline
coupled Lorentzians 14 & 447.81 & 30.79 &   0.29 & 0 && $Fe-O,\ (FeO_6)$, stretch \\
\cline{2-7}
& 520.16 &  24.38   & 0.16 && -0.26 & $Fe-O,\ (FeO_6)$, stretch \\
\hline
coupled Lorentzians 15 & 538.47 &   22.55 & 0.56 & 0 && $\nu_3\ (Fe/Mg)O_4$, stretch \\
\cline{2-7}
& 582.43    & 20.26 &   0.39 && -6.00 & $\nu_3\ (Mg/Fe)O_4$, stretch \\
\hline\hline
\multicolumn{7}{l}{$ \Delta\epsilon(0)=\epsilon_\infty+\sum \limits_{j=1}^{22} \Delta\epsilon_j=18.32 $}\\
\noalign{\smallskip}\hline
\end{tabular}
\end{table}

\subsection{Low temperature effects}
\label{sec: LowTemp}
\par We have performed measurements of the THz-IR spectra of $Ba_2Mg_2Fe_{12}O_{22}$ down to liquid helium temperatures. Figure 3 shows corresponding spectra of $\epsilon'$ and $\epsilon''$ at room temperature and at $10\ K$. Two features have to be pointed
out.
\par Firstly, the low frequency (around $15\ cm^{-1}$) value of $\epsilon'$ does not change significantly while cooling down, with the total dielectric contribution of all higher frequency absorption bands of $\Delta \epsilon \approx 18$. This means that there are no significant structural changes in the crystal lattice of $Ba_2Mg_2Fe_{12}O_{22}$ while cooling down except trivial narrowing of the bands and a certain shift of the line around $130\ cm^{-1}$ (figure 4b).

\begin{table}
\caption{Factor-group analysis of mechanical representation for $Ba_2Mg_2Fe_{12}O_{22}$ hexaferrite. In the first column, the number at the atomic symbol represents the sublattice type.}
\label{tab:2}
\begin{tabular}{|lllll|}
\hline\noalign{\smallskip}
Atom & Wyck.pos. & Site symmetry & Vibration representation & Atomic environment \\
\noalign{\smallskip}\hline\noalign{\smallskip}
O1 &    18h &   $m[C_s(6)]$ & $2A_{1g}+ A_{1u} + A_{2g} + 2A_{2u}+3E_g +3E_u$ & coplanar triangle $Fe_3$ \\
Fe2 & 18h & $m[C_s(6)]$ &   $2A_{1g}+ A_{1u} + A_{2g} + 2A_{2u}+3E_g +3E_u$ &   octahedron $O_6$ \\
O3 & 18h &  $m[C_s(6)]$ &   $2A_{1g}+ A_{1u} + A_{2g} + 2A_{2u}+3E_g +3E_u$ &   non-coplanar triangle $Fe_3$\\
O4 & 18h &  $m[C_s(6)]$ &   $2A_{1g}+ A_{1u} + A_{2g} + 2A_{2u}+3E_g +3E_u$ & tetrahedron $MgFe_3$\\
O5 &    6c &    $3m[C_{3v}(2)]$ &   $A_{1g}+ A_{2u}+E_g +E_u$ & tetrahedron $Fe_4$\\
M6 &    6c &    $3m[C_{3v}(2)]$ & $A_{1g}+ A_{2u}+E_g +E_u$ &   tetrahedron $O_4$\\
Ba7 &   6c &    $3m[C_{3v}(2)]$ &   $A_{1g}+ A_{2u}+E_g +E_u$ & 9-vertex polyhedron $O_9$\\
O8 &    6c &    $3m[C_{3v}(2)]$ &   $A_{1g}+ A_{2u}+E_g +E_u$ & tetrahedron $MgFe_3$\\
M9 &    6c &    $3m[C_{3v}(2)]$ &   $A_{1g}+ A_{2u}+E_g +E_u$ & tetrahedron $O_4$\\
Fe10 &  6c &    $3m[C_{3v}(2)]$ &   $A_{1g}+ A_{2u}+E_g +E_u$ & octahedron $O_6$\\
Fe11 &  3b &    $-3m[D_{3d}(1)]$ & $A_{2u}+E_u$ &   octahedron $O_6$\\
Fe12 &  3a &    $-3m[D_{3d}(1)]$ & $A_{2u}+E_u$ &   octahedron $O_6$\\
\hline
\multicolumn{5}{l}{$M6 = 0.75Fe + 0.25Mg; M9 = 0.75Mg + 0.25Fe$}\\
\hline\hline
\multicolumn{5}{l}{$\Gamma_{mech}= 14A_{1g}[Raman:(\alpha_{xx}+\alpha_{yy},\alpha{zz})] + 4A_{1u} (silent) + 4A_{2g} [R_z] + 16A_{2u} (IR:\ z) + $}\\
\multicolumn{5}{l}{$+18E_g [(Rz,Rz);Raman: (\alpha_{xx}-\alpha_{yy},\alpha_{xy}); (\alpha_{xz},\alpha{yz})]+ 20E_u [IR: x,y].\ \Gamma_{acoustic}= A_{2u}+E_u. $}\\
\multicolumn{5}{l}{$\Gamma_{mech}= 14A_{1g}[Raman:(\alpha_{xx}+\alpha_{yy},\alpha{zz})] + 4A_{1u} (silent) + 4A_{2g} [R_z] + 15A_{2u} (IR:\ z) + $}\\
\multicolumn{5}{l}{$+18E_g [(Rz,Rz);Raman: (\alpha_{xx}-\alpha_{yy},\alpha_{xy}); (\alpha_{xz},\alpha{yz})]+ 19E_u [IR: x,y].$}\\
\noalign{\smallskip}\hline
\end{tabular}
\end{table}

\par Secondly, we were able to register for the first time several new absorption lines appearing below $195\ K$ and below $50\ K$. These two specific temperatures correspond to the change in the magnetic structure of $Ba_2Mg_2Fe_{12}O_{22}$ \cite{13}. The new lines are indicated by arrows in figure 3 and shown in more details on expanded scales in figure 4.
\par At the temperature of $195\ K$, the configuration of the internal magnetic structure of $Ba_2Mg_2Fe_{12}O_{22}$ changes from collinear ferrimagnetic at $T > 195\ K$ (figure 1a) to an incommensurate proper screw spin-ordered at $50 < T < 195\ K$ \cite{13} with the wavevector $k_0 \parallel [001]$. This magnetic transition leads to an appearance of three new lines at $85\ cm^{-1}$, $140\ cm^{-1}$ and $232\ cm^{-1}$ (figures 3, 4). We believe that these modes have magnetic origin and are related to an excitation of optically active magnetic polaritons \cite{35,36}. To check our suggestion, optical measurements in magnetic field are in progress.
\par Another excitation was discovered in $Ba_2Mg_2Fe_{12}O_{22}$
at the lowest frequencies, below $40-50\ cm^{-1}$. As shown in
figure 4c, there is a broad wing below $80\ cm^{-1}$ in the
$\epsilon''$ spectrum at $100\ K$ that transforms into a more
pronounced absorption band at $10\ K$. In \cite{11} by using the
technique of THz time-domain spectroscopy, an absorption band has
been detected in $Ba_2Mg_2Fe_{12}O_{22}$ at $22\ cm^{-1}$ below
$50\ K$ (where the magnetic structure has changed to a
longitudinal conical \cite{13}) and has been assigned to an
electromagnon excitation. Although this line was observed in the
geometry $ \vec{E} \parallel c$, we assume that the transition
around $50\ K$, that leads to an activation of an electromagnon
for $\vec{E}\parallel c$, is accompanied by certain distortions in
the crystal lattice of $Ba_2Mg_2Fe_{12}O_{22}$ which break the
selection rules and enable excitation of the mode also for
polarization $\vec{E}\perp c$.

\begin{figure}
\centering
\resizebox{0.6\columnwidth}{!}{
%\vspace{6cm}
\includegraphics{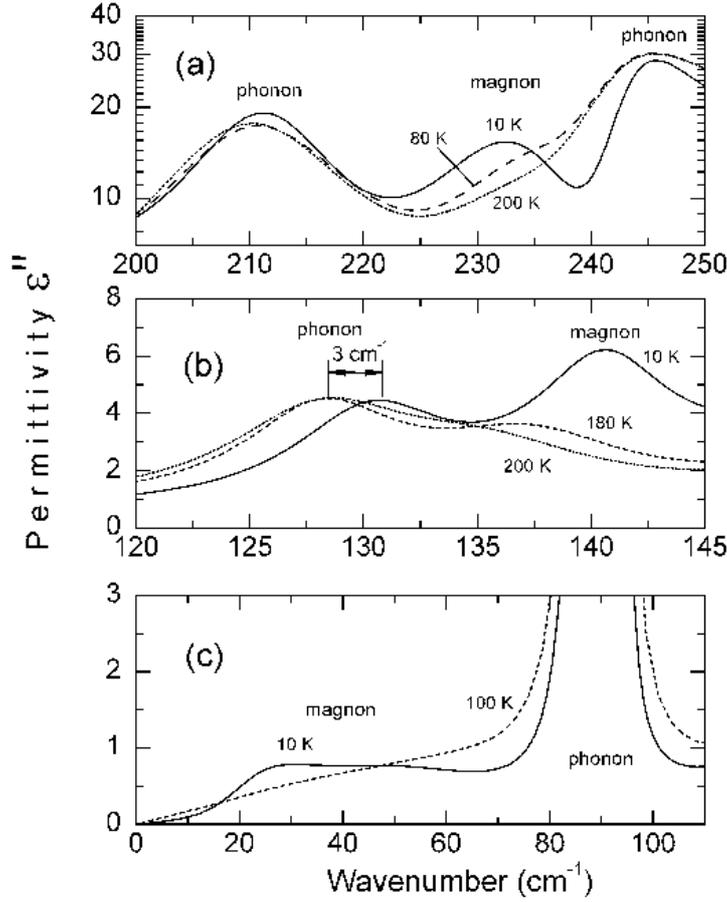}
} \caption{Spectra of imaginary part of dielectric permittivity of
$Ba_2Mg_2Fe_{12}O_{22}$ single crystal at different temperatures,
for polarization $\vec{E}\perp c$. The spectra demonstrate
appearance of new absorption lines below the magnetic phase
transition temperature $T\approx195\ K$. } \label{fig:4}
\end{figure}

\par According to the spin currents model \cite{37} (or the inverse Dzyaloshinskii-Moriya model), the non-collinear spin ordering (figure 1d) can produce a local polarization $\vec{p}_{i,j} = A\vec{e}_{i,j}\times(\vec{S}_i\times\vec{S}_j)$, that is vanishing on a macroscopic scale. Here the constant $A$ depends mainly on the spin-orbit and the exchange interactions. The neighboring spins $\vec{S}_i$ and $\vec{S}_j$, that connect sites $i$ and $j$, are aligned along the unit vector $\vec{e}_{i,j}=\vec{k}_0/\mid\vec{k}_0\mid$. Within this model, in order to create the ferroelectric polarization, the alignment of the cone axis should deviate from the direction $\vec{k}_0$. (It is important that such changes should happen, on the microstructural level, to produce the value of A not equal to zero). The $Ba_2Mg_2Fe_{12}O_{22}$ crystal is composed of an alternating stacking of the T and S blocks along the $c$ axis, as depicted in figure 1a. Within these blocks, the magnetic moments on $Fe$ ions are collinearly aligned and ferrimagnetically coupled. Along with that, the magnetic moments between the adjacent T and S blocks align noncollinearly (figure 5). The resultant magnetic structure of $Ba_2Mg_2Fe_{12}O_{22}$ in the absence of magnetic field is a noncollinear screw spiral magnetic structure. This can lead to local distortions of magnetic order and to nonzero values of A. When the temperature is lowered, the exchange interactions change their amplitudes (in particular, on the boundaries between the T and the S blocks), probably due to spin-phonon interaction with the deformation $Fe-O-Fe$ modes involved (figure 5a). It is this fact that explains the evolution of shape of the spectral response presented in figure
4.

\begin{figure}
\centering
\resizebox{0.4\columnwidth}{!}{
%\vspace{6cm}
\includegraphics{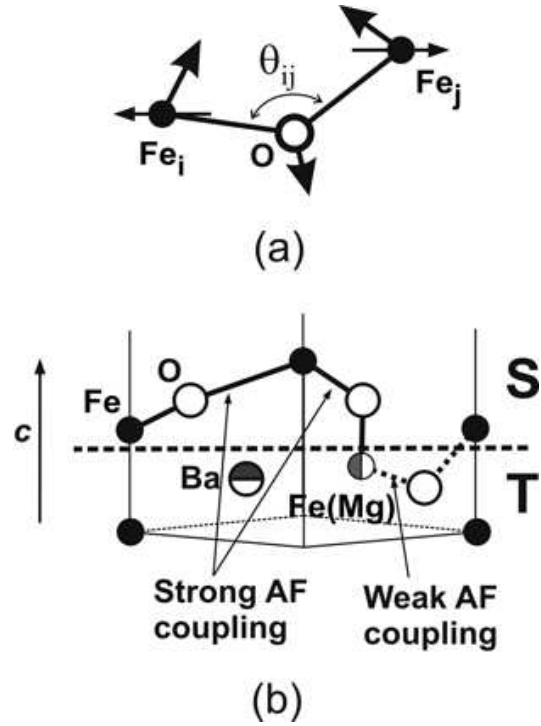}
}
\caption{Displacements corresponding to the bending mode and superexchange interactions in $Ba_2Mg_2Fe_{12}O_{22}$.}
\label{fig:5}
\end{figure}

\section{Summary}
\label{sec:summary} By using the terahertz frequency-domain and
the infrared Fourier spectrometers, the spectra of the real and
the imaginary parts of the dielectric permittivity have been
measured of a single crystalline $Y$-type hexaferrite
\\ $Ba_2Mg_2Fe_{12}O_{22}$, at frequencies $3\ cm^{-1}$ to $4000\
cm^{-1}$ in the temperature interval from $5\ K$ to $300\ K$ and
for polarization $\vec{E}\perp c$.
\par New absorption lines are discovered below the temperatures of $195\ K$ and $50\ K$, where the internal magnetic field configuration in $Ba_2Mg_2Fe_{12}O_{22}$ changes. The origin of these lines is assigned to optical polaritons. Their spectral weight is nonzero in the phase with longitudinal spin polarization ($T < 50\ K$) testifying the influence of the spin-lattice interaction (that involves the deformational $Fe-O-Fe$ modes) on the superexchange interaction at the "junction" between the S and the T
structural blocks.
\par The parameters of all
nineteen (allowed by the $R 3 m$ symmetry) phonon $E_u$ modes are
determined. Their assignment to the vibrational modes of the
crystal lattice fragments is performed.
\par It is found that the total contribution to the static dielectric permittivity of the polar phonon modes is practically independent on the temperature in the range $10\ K-300\ K$, implying no significant structural changes in the crystal lattice of $Ba_2Mg_2Fe_{12}O_{22}$.

\section{Acknowledgements}
\label{sec:Acknowledgements}
Funding from Russian Foundation for Basic Research (project 09-02-00280-a) is acknowledged.

\end{document}